\documentclass[10pt,twocolumn,prl,superscriptaddress,english]{revtex4-1}
\usepackage[T1]{fontenc}
\usepackage[latin9]{inputenc}
\usepackage{amsmath}
\usepackage{amssymb}
\usepackage{graphicx}

\makeatletter

\usepackage{babel}

\makeatother

\usepackage{babel}

\usepackage{color}

\newcommand{\bC}{{\mathbb C}}

\begin{document}

\title{Algebraic Geometrization of the Kuramoto Model: Equilibria and Stability Analysis}

\author{Dhagash Mehta}
\email{dmehta@nd.edu}
\affiliation{Dept.~of Applied and Computational Mathematics and Statistics, University of Notre Dame, Notre Dame, IN 46556, USA.}

\author{Noah S. Daleo}
\email{nsdaleo@ncsu.edu}
\affiliation{Dept.~of Mathematics, North Carolina State University, Raleigh, NC 27695, USA.}

\author{Florian D\"orfler}
\email{{dorfler@ethz.ch}}
\affiliation{Automatic Control Laboratory, Swiss Federal Institute of Technology (ETH) Z\"urich, 8092 Z\"urich, Switzerland.}

\author{Jonathan D.~Hauenstein}
\email{hauenstein@nd.edu}
\affiliation{Dept.~of Applied and Computational Mathematics and Statistics, University of Notre Dame, Notre Dame, IN 46556, USA.}

\begin{abstract}
\noindent 
Finding equilibria of the finite size Kuramoto model amounts to 
solving a nonlinear system of equations, which is an important yet challenging problem. 
We translate this into an algebraic geometry problem and 
use numerical methods to find all of the equilibria for various 
choices of coupling constants $K$, natural frequencies, and on different graphs.
We note that for even modest sizes ($N\sim 10-20$), 
the number of equilibria is already more than 100,000.
We analyze the stability of each computed equilibrium as well as the configuration of angles. 
Our exploration of the equilibrium landscape leads to unexpected and possibly surprising results
including non-monotonicity in the number of equilibria, a predictable pattern in the 
indices of equilibria, counter-examples to popular conjectures, multi-stable equilibrium landscapes, scenarios with only unstable equilibria, and multiple distinct extrema in the stable equilibrium~distribution as a function of the number of cycles~in~the~graph.
\end{abstract}

\maketitle

\noindent\textit{Introduction.~-~}
The Kuramoto model is a fascinating model proposed in 1975
to study synchronization phenomena \cite{kuramoto1975self}
that has gained attention from various scientific communities, including
biology, chemistry, physics, and electrical engineering,
due to its applicability. 
This model has been used to study various phenomena including
neural networks, chemical oscillators, Josephson junctions and laser arrays, 
power grids, particle coordination, 
spin glass models, and rhythmic applause~\cite{acebron2005kuramoto,strogatz2000kuramoto,FD-FB:13b}.
  
The Kuramoto model is defined as a system of autonomous ordinary differential equations as:
\begin{equation}\label{Eq:KuramotoStandard}
\frac{d\theta_{i}}{d t}=\omega_{i} - \frac{K}{N}\sum_{j=1}^{N}a_{i,j}\sin(\theta_{i}-\theta_{j}),\mbox{~for~}i=1,...,N,
\end{equation}
where $K$ is the coupling strength, $N$ is the number of oscillators, 
$\Omega=(\omega_{1},\dots,\omega_{N})$ is the vector of intrinsic natural frequencies, and 
$a_{i,j} \in \{0,1\}$ is the $(i,j)$th element of the 
adjacency matrix of the coupling graph.
The natural frequencies $\omega_i$ indicate how the system oscillates in 
the absence of any dissipation or exogenous forces.

The equilibrium conditions are $\frac{d\theta_{i}}{d t}=0$ for all $i$.
This system of equations has an $O(2)$ freedom, i.e.,
for any \mbox{$\alpha\in(-\pi,\pi]$}, the equations are invariant 
under replacing all $\theta_i$ with $\theta_i+\alpha$.
This rotational symmetry leads to a continua of 
equilibria. To remove this $O(2)$ freedom resulting in 
finitely many equilibria, we fix one of the angles, say, 
$\theta_{N}=0$, and remove the equation 
$\frac{d \theta_{N}}{d t}=0$ from the system. The remaining 
system consists of $N-1$ nonlinear equations in $N-1$ angles.

Provided the coupling strength $K$ is strong enough, 
the oscillators will synchronize as $t\rightarrow\infty$. 
In this setup, a critical coupling $K_c(N)$ exists at which 
the number of stable equilibria switches from $0$ to a nonzero value. 
In the special case of $N\rightarrow\infty$ and long-range (all-to-all) coupling $a_{ij} = 1$, one may analytically 
compute $K_c(N)$.  
However, for the finite size Kuramoto model, such an analysis may turn out to be very difficult. 
In particular, finding all equilibria, analyzing stability, and 
finding $K_{c}(N)$ is known to be
prohibitively difficult for a finite but large oscillator population. 

We point out that the equilibria of system (\ref{Eq:KuramotoStandard}) can also be viewed as the stationary points of the potential energy landscape drawn by the 
mean-field XY model with an exogeneous perturbation term:
\begin{equation}
\label{eq: potential}
 V(\theta) = \frac{K}{2N} \sum_{i,j=1}^{N} a_{i,j}(1-\cos (\theta_{i} - \theta_{j})) - \sum_{i=1}^{N}\omega_{i} \theta_{i},
\end{equation}
whose gradient 
reproduces the right-hand side of equation~\eqref{Eq:KuramotoStandard}. Hence, in the following,
we use the words equilibria and stationary points interchangeably.

All stationary points of the finite $N$ mean-field XY model (i.e., the Kuramoto model with homogeneous frequencies)
were identified in Ref.~\cite{Casetti:June2003:0022-4715:1091}. 
Building on Ref.~\cite{Casetti:June2003:0022-4715:1091},
all stationary points of the one-dimensional 
nearest-neighbour $XY$ model (i.e.,~the Kuramoto model with local coupling) 
for any given $N$ with either periodic or anti-periodic boundary conditions
have been found~\cite{Mehta:2009,Mehta:2010pe,vonSmekal:2007ns,vonSmekal:2008es}.

Using these solutions, a class of stationary points of the 2-dimensional 
nearest-neighbour $XY$ model \cite{Nerattini:2012pi} and the $XY$ model
with long-range interactions \cite{kastner2011stationary} were built and analysed (see also \cite{Hughes:2012hg,Mehta:2014jla}). 
In Ref.~\cite{Mehta:2009,Mehta:2010pe,Mehta:2009zv,Hughes:2012hg}, all of the stationary points for small lattices were found 
using algebraic geometry methods. Bounds on the number of equilibria  \cite{JB-CIB:82} as well as some counterintuitive 
examples to plausible conjectures \cite{AA-SS-VP:81} have been reported for the same model in the domain of power systems. 

In Ref.~\cite{Casetti:June2003:0022-4715:1091} for the finite $N$ mean-field $XY$ model, and
in Refs.~\cite{Mehta:2009,Nerattini:2012pi} for the nearest neighbour $XY$ models, it was 
shown that there were exponentially many isolated stationary solutions 
as $N$ increases.  Moreover, even after breaking the global $O(2)$ 
symmetry, there were continuous solutions at the maximum value of the 
energy which were independently observed and 
termed as {\em incoherent manifolds} in Ref.~\cite{strogatz1991stability}.

In Refs.~\cite{verwoerd2009computing,verwoerd2008global,dorfler2011critical}, 
necessary and sufficient conditions are given for fixed points to 
exist for the finite 
size Kuramoto model for complete and bipartite graphs, and explicit upper 
and lower bounds of $K_c(N)$ for these systems were also computed, followed 
by providing an algorithm to compute $K_c(N)$. For the complete graph, similar results were presented in \cite{DA-JAR:04,REM-SHS:05}, where it was additionally shown that there is exactly one single stable equilibrium for $K > K_c(N)$. 
In Ref. \cite{FD-MC-FB:11v-pnas}, an analogous result was shown for acyclic graphs, short cycles, and complete graphs as well as combinations thereof.
In the case of 
homogeneous natural frequencies networks with sufficiently high 
nodal degrees, the only stable fixed point is known to be the phase-synchronized solution \cite{taylor2012there}.
In Ref.~\cite{ochab2009synchronization}, 
all of the stable synchronized 
states were classified for the one-dimensional Kuramoto model on a ring 
graph with random natural frequencies in addition to computing a lower 
bound on $K_c(N)$ (see also Ref.~\cite{ermentrout1985behavior,strogatz1988phase, tilles2011multistable}). 

\noindent\textit{Algebraic Geometry Interpretation and Setup.~-~}
We initiate an approach to study the Kuramoto model by 
using an algebraic geometry interpretation of the 
equilibria and studying synchronization. 
Upon fixing $\theta_N = 0$, 
we use the identity 
$\sin(x-y)=\sin x \cos y - \sin y \cos x$
and substitute $s_i = \sin\theta_i$ and $c_i = \cos\theta_i$
to transform the $N-1$ equations into polynomials
that are coupled by the Pythagorean identity
$s_i^2+c_i^2=1$.
This results in a system of $2(N-1)$ polynomials
in $2(N-1)$ variables:
\begin{equation}\label{Eq:PolySystem}
\begin{array}{l}
0 = \omega_i+\frac{K}{N}\sum_{j=1}^{N} a_{i,j} \left( s_i c_j - s_j c_i \right) \\
0 = s_i^2 + c_i^2 - 1
\end{array}
\end{equation}

Given a fixed integer $N$, our first goal is to find all real 
solutions of polynomial system (\ref{Eq:PolySystem}) for a given choice of 
$\Omega$, $K$, and adjacency matrix $A=[a_{ij}]$. 
We accomplish this using a technique from numerical algebraic geometry 
via parameter homotopies, which we briefly review below.
Once the system is solved, we 
determine which solutions (if any) are stable steady-state 
solutions by analyzing the eigenvalues of the Jacobian at each real solution.

\noindent\textit{Numerical Algebraic Geometry Methods.~-~}
For a fixed $N$, we can interpret \eqref{Eq:PolySystem} 
as the system $F_N(s,c;K,\Omega,A) = 0$ with variables
$s$ and $c$ and parameters $K$, $\Omega$, and $A$.
This leads to using parameter homotopies \cite{CoeffParam}
for solving, which is a two-step process.
First, in the {\em ab initio} phase, one computes all solutions for a 
sufficiently random set of parameters.  
Then, in the parameter homotopy phase, one solves for given parameters 
by deforming from the parameters selected in the {\em ab initio} phase.
The {\em ab initio} phase is performed once, while the parameter homotopy phase
is performed for each set of parameters.
The following provides a short introduction with
more details provided in \cite[Ch.~7]{SW05} and \cite[Ch.~6]{BHSW13}.

This process is a generalization of other standard homotopy methods,
such as a \textit{total degree homotopy} or \textit{multi-homogeneous homotopy}
frequently used, e.g.,~\cite{SW05, BHSW13,Mehta:2011wj,Maniatis:2012ex,Kastner:2011zz,Mehta:2012wk,Mehta:2012qr,Greene:2013ida,Mehta:2013fza,MartinezPedrera:2012rs,He:2013yk}.
The method has also recently used in solving the power-flow systems \cite{konsta,chandra_mehta14}.
For example, consider solving a system of $m$ polynomial equations in $m$ variables, $f(x)=(f_1(x),\dots,f_m (x))=0$ where $x=(x_1,\dots, x_m)$.
For a total degree homotopy, the space of interest 
is the space of all polynomial systems $g(x)$ of $m$ polynomials in $m$ variables
where $d_i := \deg f_i = \deg g_i$.  Since a random element in this space
has the B\'ezout number of roots, namely $\prod_{i=1}^m d_i$,
the system $G = (G_1,\dots,G_m)=\textbf{0}$ where $G_i(x) = x_i^{d_i} - 1$
is sufficiently random in this space.  

In this case, the {\em ab initio} phase for solving $G = 0$ is trivial.
Then, in the parameter homotopy phase, one can consider the
homotopy $H(x,t) = (1-t)f(x) + \gamma t G(x)$ for a random $\gamma\in\bC$.
Now, the solutions of $H(x,t) = 0$ at $t=1$ are known 
and we want to compute the solutions of $H(x,0) = 0$.  
Using a numerical predictor-corrector method implemented 
in {\tt Bertini} \cite{Bertini}, we track each of the solution paths from $t=1$ to $t=0$. 
The number~$\gamma$ ensures that such path tracking will obtain \textit{all} of the isolated complex solutions of $f(x)=0$, from which the real and nonreal solutions can be identified. 

Now, returning to our case $F_N$, suppose, for simplicity, 
that we want to study the behavior of the solutions 
of $F_N = 0$ as a function of $K$, that is, we also fix $\Omega$ and $A$.  
In the {\em ab initio} phase, we pick a random $K = K_0\in\bC$ 
and compute the solutions $S_0$ of $F_N(s,c;K_0,\Omega,A) = 0$
using a total degree homotopy described above via {\tt Bertini}.

In the parameter homotopy phase, we consider solving for
various choices of parameters $K$, e.g., $K=2,\ldots,100$.  
For each choice of $K$, we simply track the solution paths of
$F_N(s,c;K\cdot (1-t) + t\cdot K_0,\Omega,A) = 0$ starting
at $t = 1$ with solutions $S_0$.  
Each of these computations is relatively inexpensive, thereby making 
it practical to compute the solutions at hundreds of different 
choices for $K$.

The resulting solutions are sorted based on whether they are real or not, and their stability
is investigated as follows. From the  sine $s_i$ and cosine $c_i$ values, we compute
the corresponding angle $\theta_i$.  The eigenvalues of the 
Jacobian of \eqref{Eq:KuramotoStandard} are used 
to analyze stability.  The {\em index} is the number of positive
eigenvalues so that real solutions with index zero correspond
to stable steady states.



\begin{figure}[ht]
  \centering
  \includegraphics[width=0.5\textwidth]{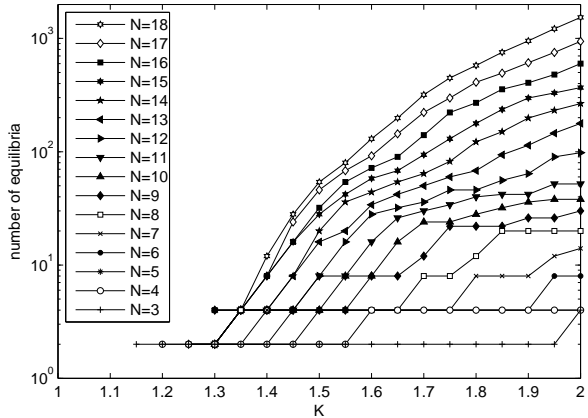}
  \caption{Number of equilibria for the case of equidistant natural frequencies on the complete graph at $1\leq K \leq 2$.}
\label{figure:NumEquilibriaSmallK}
\end{figure}

\begin{figure}[ht]
  \centering
  \includegraphics[width=0.5\textwidth]{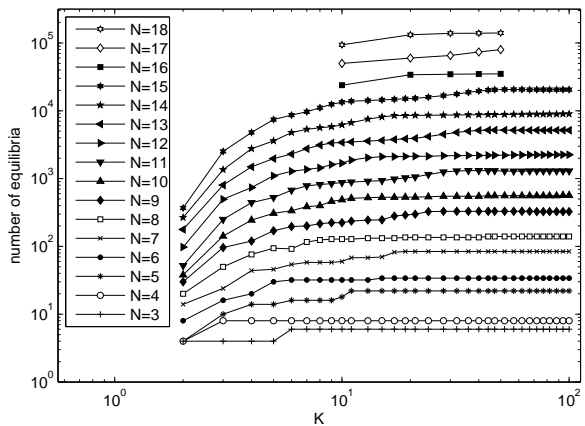}
  \caption{Number of equilibria for the case of equidistant natural frequencies on the complete graph at $2\leq K \leq 100$.}
\label{figure:NumEquilibriaLargeK}
\end{figure}

\noindent\textit{Results for the complete graph.~-~}
We start with the most prominent and well studied case corresponding
to the complete graph, namely $a_{i,j} = 1$.
We select $\Omega=(\omega_1\ldots\omega_N)$ to be $N$ equidistant numbers, 
namely $\omega_i=-1 + (2i -1)/N$.
Figures~\ref{figure:NumEquilibriaSmallK} and~\ref{figure:NumEquilibriaLargeK}
summarize the number of real-valued solutions to the 
polynomial system~\eqref{Eq:PolySystem} for $N=3,\ldots,18$ at various values of $K$.
%
In general, we see that the number of distinct real solutions tends to grow as $K$ 
increases.  However, there are some surprising exceptions to this expected monotinicity property, e.g., the $N=9$ case has
$328$ real solutions at $K=60$ and $326$ real solutions at $K=70$.  Our 
results suggest that eventually the number of real solutions stabilizes 
to a constant for each $N$. 
For $N=18$, we computed 140,356 real solutions when $K=50$; 
computing these without our polynomial formulation would be very difficult.

\begin{figure}[ht]
\centering
\includegraphics[width=0.5\textwidth]{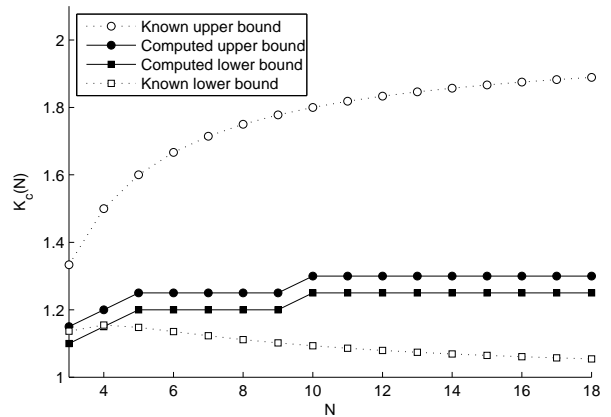}
\caption{Bounds for $K_c(N)$ in the case of equidistant natural frequencies on the complete graph.  Known explicit bounds from \cite[Corollary 6.7]{FD-FB:13b} are shown for comparison.}
\label{figure:EquidistantK_c}
\end{figure}

Once the real solutions are found, we turn our attention to determining 
which ones are stable.  Via Jacobian analysis we find that there 
is exactly a single stable steady state solution for $N=3,\ldots,18$ at each 
$K$ value tested, with the exception of small values of $K$ which 
result in no real solutions.  This result is consistent with theoretic findings for the complete graph \cite{DA-JAR:04,REM-SHS:05,dorfler2011critical,verwoerd2008global}.
{Next, for each $N$ we use the results of our stability analysis 
to determine numerical upper and lower bounds on $K_c(N)$ as follows. 
We acquire an upper bound on $K_c(N)$ by determining 
the smallest tested value of $K$ for which a real steady state is 
found.  Similarly, we find a lower bound on $K_c(N)$ by determining 
the largest tested value of $K$ for which no real steady states 
are found.}
These results are presented in Figure \ref{figure:EquidistantK_c}. 
We verify that these bounds for $K_c(N)$ are consistent 
with the known explicit bounds from \cite[Corollary 6.7]{FD-FB:13b}. 
For $N \in \{3,4\}$, our computed lower bound is less than the 
explicit lower bound due to the coarse resolution of tested numeric values for $K_c(N)$.


In Ref.~\cite{AA-SS-VP:81}, it is conjectured that if there is a stable 
equilibrium, then there is a stable equilibrium satisfying 
$|\theta_{i}-\theta_{j}| < \pi/2$ for all neighbors $\{i,j\}$. 
We checked this conjecture and we quickly found 
counterexamples. 
For $N=3$ and $K=1.15$, there is one stable 
equilibria at 
$(\theta_1,\theta_2,\theta_3)\approx (1.6820,0.8410,0)$.
Since $\theta_1$ and~$\theta_3$ are coupled with
$|\theta_{1}-\theta_{3}| > \pi/2$, we reject the conjecture.

For larger $N$, the unique stable equilibria can have angles
even more spread out, such as
$(\theta_{1},\ldots,\theta_{15})\approx (2.0981,1.8867,\ldots,0.2114,0)$ 
when $N=15$ and $K=1.3$.
These interesting numerical observations prompt us to analytically reject the conjecture in \cite{AA-SS-VP:81}: in \cite{FD-FB:13b} it is shown that the worst-case equilibrium configuration for $K=K_{c}$ and $n>3$ corresponds to a tripolar distribution of all angles spanning a half-circle. From \cite{DA-JAR:04,REM-SHS:05}, we know that this is the only stable equilibrium. These two arguments suffice to reject the conjecture in \cite{AA-SS-VP:81}.

\begin{figure}[ht]
\centering
\includegraphics[width=0.5\textwidth]{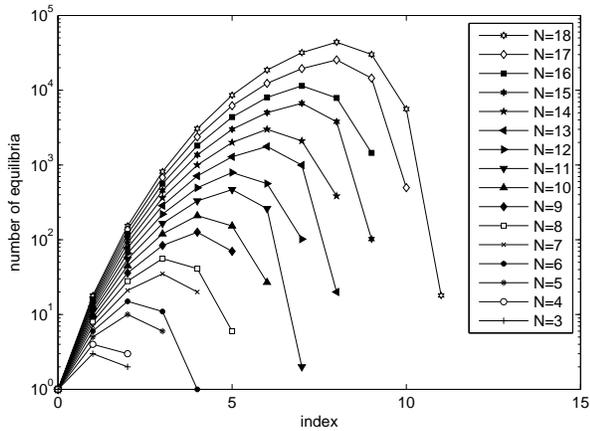}
\caption{Number of equilibria with given index for equidistant natural frequencies and a complete graph with $K=100$.}
\label{figure:IndexPlotN3To15}
\end{figure}

As mentioned earlier, our stability analysis is 
based on computing the index corresponding to each real solution.  
Figure~\ref{figure:IndexPlotN3To15} shows a histogram of 
these values for the case $K=100$.
We observe that when $j$ is small enough relative to $N$, the number 
of real solutions with index $j$ is exactly $\binom{N}{j}$.  
In particular, 
for $3\leq N\leq 4$ this behavior occurs for $j\leq 1$ and,
for $5\leq N\leq 18$, this behavior occurs for 
$j\leq \lceil{\frac{2}{5}(N-1)}\rceil$.
Based on these results, we conjecture that this phenomenon occurs in general: for equidistant natural frequencies, 
if $0 \!\leq\! j \!\ll\! N$ and $K\gg 0$, then the number of equilibria with index $j$ is 
expected to be $\binom{N}{j}$.
This conjecture can be used when $N$ may be 
too large to compute all solutions.  
For example, we expect 75,287,520 real equilibria of 
index $5$ for $N = 100$.

\noindent\textit{Results for cyclic graphs.~-~}
Although our investigation initially focused on the complete graph being the most well studied case, we also performed a preliminary 
analysis of the coupling arrangement defined by an undirected cyclic graph. 
Cyclic graphs are known for having multi-stable equilibria \cite{FD-FB:13b}, entirely unstable equilibrium landscapes \cite{AA-SS-VP:81}, and thus a quite distinct behavior from acyclic or complete graphs \cite{FD-MC-FB:11v-pnas,wiley2006size,ermentrout1985behavior}.

\begin{figure}[ht]
\centering
\includegraphics[width=0.5\textwidth]{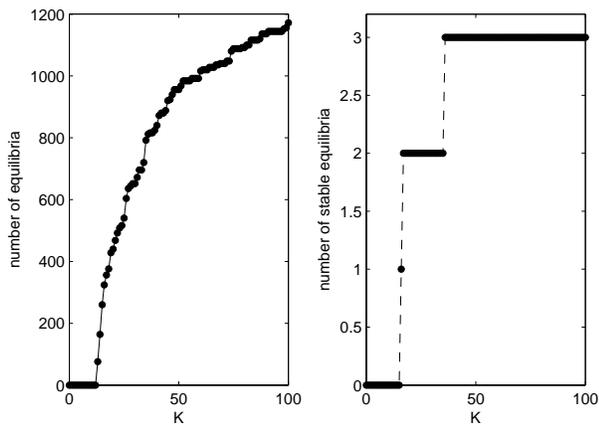}
\caption{Number of equilibria and number of stable equilibria for a cyclic coupling arrangement when $N=10$.}
\label{figure:CyclicGraph}
\end{figure}

For $N=10$, we used the same equidistant natural frequencies mentioned 
earlier with Figure \ref{figure:CyclicGraph} showing the number of equilibria and 
the number of stable equilibria at each integer $K=0,\ldots,100$.

For some values of $K$, the system possesses only unstable equilibria.
In particular, when $K$ is $13$, $14$, or $15$, there are $76$, $164$, and $260$ equilibria, respectively, all of which are unstable.
Since the critical points of the potential \eqref{eq: potential} correspond to power flow in a transmission network, a very interesting technological implication of these results is that there are power demands which can be met only with unstable equilibria. 
We find one stable equilibria at $K=16$ and conclude $15\leq K_c(10)\leq 16$. 
Figure~\ref{figure:CyclicGraph} also indicates multistability for some values of $K$, 
with at most {\em three} stable equilibria for each investigated sample. 
These results are in contrast to the complete graph case, in which exactly one stable equilibria exists whenever the system has real-valued solutions. 

Next, we take a closer look at the geometric configuration of angles occurring at the stable equilibria shown in Figure \ref{figure:CyclicGraph}.
For relatively small $K$ such as $K=16,17,18$, the configuration of angles at stable equilibria shows no discernible structure.
For $K=35$ there are two stable equilibria, with one exhibiting phase sync, i.e., angles clustered around $0$, and one exhibiting a splay state, 
i.e., angles approximately uniformly distributed on $[0,2\pi)$.
For the splay state, the $\theta_i$ decrease on $[0,2\pi)$.

\begin{figure}[ht]
\centering
\includegraphics[width=0.5\textwidth]{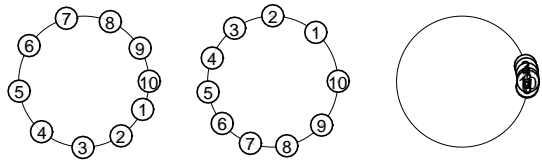}
\caption{Configuration of $\theta_1,\ldots,\theta_{10}$ for the three stable~equilibria occurring at $K=100$ for a cyclic graph with $N = 10$.}
\label{figure:CyclicConfiguration}
\end{figure}

For each $K=36,\ldots,100$ there are three stable equilibria, with each case consisting of one phase sync and two splay states.
In one of these splay states $\theta_1,\ldots,\theta_9$ are arranged in increasing order on $[0,2\pi)$, and in the other $\theta_9,\ldots,\theta_1$ are arranged in decreasing order on $[0,2\pi)$.
In Figure \ref{figure:CyclicConfiguration}, we depict the three stable equilibria at $K=100$. 
As $K$ increases, the steady state phase sync gradually becomes more tightly clustered, with the angle range decreasing from $\sim0.9432$ for $K=35$ to $\sim0.3256$ for~$K=100$. 
This is in accordance with the asymptotic result that exact phase sync is a critical point of the Kuramoto potential~\eqref{eq: potential} as $K\to \infty$ \cite{FD-FB:13b}.

\noindent\textit{Results for random graphs.~-~}
We now turn our attention to random coupling arrangements.
In these computations, we choose each $a_{i,j}$ as $0$ or $1$ according to a pre-determined probability $P$ while setting $a_{j,i}=a_{i,j}$.
In other words, $[a_{i,j}]$ is the adjacency matrix of a symmetric Erd\"{o}s-R\'enyi random graph with coupling probability~$P$, where we restrict ourselves to connected graphs.
For these numerical experiments, we use the same equidistant natural frequencies discussed\,earlier.

\begin{figure}[ht]
\centering
\includegraphics[width=0.5\textwidth]{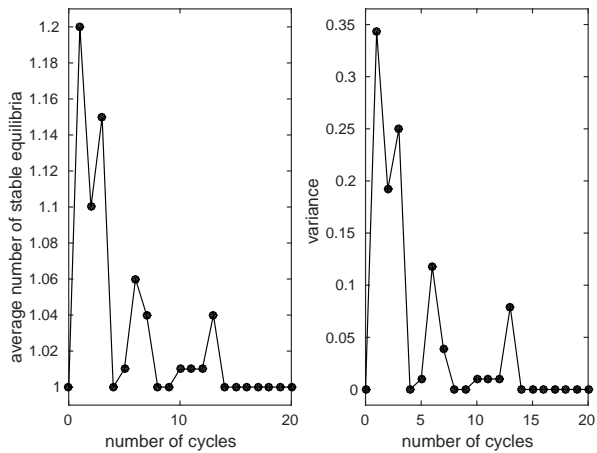}
\caption{Number of stable equilibria for random graphs according to number of cycles in the case of $N=8$ and $K=100$.}
\label{figure:RandomGraphCyclesvsNumStable}
\end{figure}

First, we fix $N=8$ and $K=100$ and investigate the number of 
stable equilibria for sparse random graphs.
For each $c=0,1,\ldots,20$, we construct $100$ random graphs having exactly $c$ cycles. 
{We achieve this by generating adjacency matrices of symmetric Erd\H{o}s-R\'enyi random graphs until we have $100$ instances of connected graphs with the desired number of cycles.}
Depending on the graph, we find $1$, $2$, or $3$ stable equilibria.
The averages of these results are shown in Figure~\ref{figure:RandomGraphCyclesvsNumStable}
showing that the sparsest (i.e., acyclic) as well as sufficiently dense graphs have exactly one stable equilibrium confirming the analytic results for these two extremal  topologies \cite{FD-MC-FB:11v-pnas,FD-FB:13b,DA-JAR:04,REM-SHS:05,taylor2012there}. Since cycles are known to display multi-stability (see Figure~\ref{figure:CyclicGraph}), it is to be expected that there is more than one stable equilibrium for small cycle numbers.  Quite surprisingly, multiple distinct extrema can be observed in the stable equilibrium distribution in Figure~\ref{figure:RandomGraphCyclesvsNumStable}.

\begin{figure}[ht]
\centering
\includegraphics[width=0.5\textwidth]{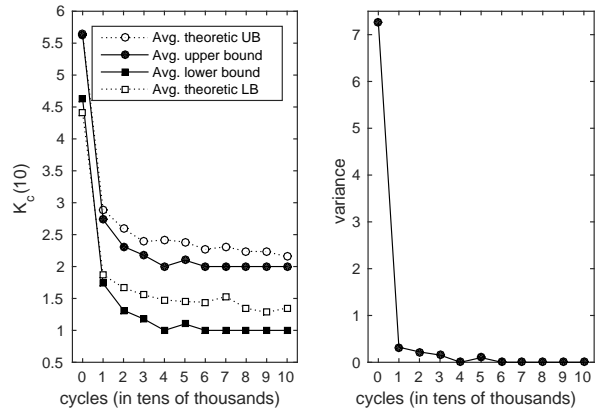}
\caption{Average bounds on $K_c(N)$ for random graphs according to number of cycles in the case of $N=10$.}
\label{figure:RandomGraphCyclesvsK_c}
\end{figure}

Next, we fix $N=10$ and investigate how $K_c(N)$ depends on the density of the graph. 
For each coupling probability $P=0.25,0.375,\ldots,0.875$, we generated $100$ random graphs. 
For each graph, we computed equilibria and determine stability at the following values of $K$: 
for $P=0.25$ we use $K=1,\ldots,20$; for $P=0.375$ we use $K=1,\ldots,15$; for $P=0.5$ we use $K=1,\ldots,10$; and for each $P=0.625,0.75,0.875$, we use $K=1,\ldots,5$.
In all cases, we find a value of $K$ such that at least one stable equilibrium occurs at $K$ while no stable equilibria occur at $K-1$, thereby allowing us to estimate bounds on $K_c(10)$. 
Figure~\ref{figure:RandomGraphCyclesvsK_c} shows these results sorted according 
to the number of cycles in the graph.

For undirected and connected graphs,
a theoretic lower and a conjectured theoretic upper bound are \cite{FD-FB:13b,FD-MC-FB:11v-pnas} 
$$N\cdot \max_i \left\{ \frac{|\omega_i|}{\textup{deg}_{i}} \right\} \leq K_c(N)\leq N\cdot \| B^T L^\dagger \omega \|_\infty,$$
where $\textup{deg}_{i} = \sum_j a_{ij}$ is the degree of node $i$, $B$ is the oriented incidence matrix, and $L$ is the network Laplacian matrix. 
These bounds are shown in Figure \ref{figure:RandomGraphCyclesvsK_c} for comparison. 
We verified that our $600$ individual results as well as the averages shown are consistent with these theoretic bounds and validate their accuracy. 
Figure~\ref{figure:RandomGraphCyclesvsK_c} indicates that the numerical upper bound is tighter than the theoretic bound on average, while the numerical lower bound tends to be weaker. 
Since the tightness of our computed bounds depends on the resolution of $K$ values tested, one could check more refined $K$ values to obtain tighter numerical bounds for particular graphs of interest.


In future work, we aim to extend our investigations to directed graphs, negative weights, 
cosine coupling, and statistical properties of random graph models, as well as verify conjectures for variations of the Kuramoto models
as studied in Ref.~\cite{wiley2006size}.

\noindent\textit{Acknowledgement.~-~}
DM would like to thank Carlo Laing and Steven Strogatz for their feedback at the initial stages of this work. 
NSD, JDH, and DM were supported by DARPA Young Faculty Award
with NSD and JDH additionally supported by NSF DMS-1262428.
FD was supported by in part by ETH Z\"urich startup funds.

\bibliographystyle{unsrt}



\end{document}